\documentstyle[12pt]{article}
\begin{document}

\newcommand{\be}{\begin{equation}}
\newcommand{\ee}{\end{equation}}
\newcommand{\kz}{|K^0\rangle}
\newcommand{\kzb}{|\bar{K^0}\rangle}
\newcommand{\ko}{|K_1\rangle}
\newcommand{\kt}{|K_2\rangle}
\newcommand{\ks}{|K_S\rangle}
\newcommand{\kl}{|K_L\rangle}
\newcommand{\bz}{|B^0\rangle}
\newcommand{\bzb}{|\bar{B^0}\rangle}
\newcommand{\bl}{|B_L\rangle}
\newcommand{\bh}{|B_H\rangle}

\baselineskip=14pt plus 0.2pt minus 0.2pt
\lineskip=14pt plus 0.2pt minus 0.2pt

\begin{flushright}
hep-ph/9509370 \\
LA-UR-95-2030 \\
\end{flushright}

\begin{center}
\Large{\bf Quantum Interference:  From Kaons to Neutrinos (with Quantum
Beats in between)}

\vspace{0.25in}

\large

\bigskip

 Michael Martin Nieto\footnote
{Email: mmn@pion.lanl.gov \\
This work is dedicated to the memory of Bernie Deutch, our
friend and colleague, whose love of physics was
exemplified by his words and by his actions.  } \\

\vspace{0.3in}

{\it
Abteilung f\"ur Quantenphysik\\
Universit\"at Ulm\\
D-89069 Ulm, GERMANY\\
\vspace{.25in}
and\\
\vspace{.25in}
Theoretical Division\\
Los Alamos National Laboratory\\
University of California\\
Los Alamos, New Mexico 87545, U.S.A.\\}

\normalsize

\vspace{0.3in}

{ABSTRACT}

\end{center}

\begin{quotation}


Using the vehicle of resolving an apparent paradox,
a discussion of quantum interference is presented.  The understanding
of a  number of different
physical phenomena can be unified, in this context.
These range from  the neutral kaon
system to massive neutrinos, not to mention quantum beats, Rydberg wave
packets, and neutron gravity.

\vspace{0.25in}

\end{quotation}

\vspace{0.3in}

\newpage

\section{The Neutral Kaon System}


One of the most  important ``modern,
quantum-interference"
phenomenon was discovered in particle physics.  The background was that,
in the 1950's,
it was observed  that  the strongly-interacting,
neutral, $K^0$ meson,  sometimes appeared to decay via the weak interaction
 into  $\pi^+ \pi^-$.
Mind you, the full quantum field theory of the $CPT$ theorem was not
 formulated until 1957 \cite{cpt}.
Even so, in 1955 Gell-Mann and Pais \cite{gm} predicted an astounding new
effect on the bsis of these experimental results.

They predicted (in terms of our present terminology) that there
must be an antiparticle to the $K^0$, called the $\bar{K^0}$, with
opposite quantum number ``strangeness" or ``hypercharge."  Further,
the origin of the decay to the $2\pi$ state is not the pure
strangeness eigenstate, but rather the coherent mixture
\be
|K_{1}\rangle = \frac{1}{\sqrt2}\bigg( |K^0\rangle
             + |\bar{K^0}\rangle \bigg) ~.
\label{k1e}
\ee
This state is  an eigenstate of $CP$ with eigenvalue $+1$.
Another prediction was that there should be an
eigenstate of $CP$ with eigenvalue $-1$.  This state would be
\begin{equation}
|K_{2}\rangle = \frac{1}{\sqrt2}\bigg(|K^0\rangle
             - |\bar{K^0}\rangle\bigg)~.
\label{k2e}
\end{equation}
The $K_2$ would have a slightly different mass than the $K_1$, and
have a much longer lifetime.  Finally, because of this superposition,
a beam of particles that was originally composed of
$K^0$'s (or $\bar{K^0}$'s)  would
evolve in time in an interference mode,  oscillating between the two
decaying species.  In particular, inverting Eqs. (\ref{k1e}) and (\ref{k2e}),
inserting the time-dependence, and using the proper time,
\be
|K^0(\tau)\rangle =  \frac{1}{\sqrt2}
             \bigg[\exp[-(im_S +\Gamma_S/2)\tau] \ko
          +   \exp[-(im_L +\Gamma_L/2)\tau] \kt \bigg] ~,
\ee
where the $m$'s are the masses and the $\Gamma$'s are the decay widths
of the shorter- and longer-lived particles.
Similarly for the $\bar{K^0}$'s.

Eventually, all of these predictions were found to be true.  The first
verification was the discovery
of the long-lived $K_2$, which decays into a $3\pi$ state \cite{3pi}.
But there was a problem in verifying the interference.  The $K_1$'s
and the ${K_2}$'s did not decay into the same final states.
Therefore, one could not measure  $|\langle F|K^0(\tau)\rangle|^2$,
for some final state $F$.

Pais and Piccioni overcame this problem
\cite{pais} when they realized
one could go back and use the strong interactions to remix the particles.

Since the $K_1$'s and $K_2$'s are time-dependent linear combinations
of the $K^0$'s and $\bar{K^0}$'s, the reverse is also true.  Thus,
at any given time, $|K^0(\tau)\rangle$ is not only a linear combination
of $\ko$ and $\kt$, it is also a linear combination of $\kz$ and $\kzb$.
If this state is then sent into a ``regenerator,"  say a slab of copper,
the $\kz$  forward scattering amplitude, $f(0)$,
will be  different and smaller than the $\kzb$
forward scattering amplitude, $\bar{f}(0)$.
Thus, if the size of the regenerator is varied, the outcoming decay
particles will be varying  different relative superpositions
of $\ko$ and $\kt$.  After a thorough analysis
by Good \cite{good},
experiments were done, and the effect was seen \cite{regen1,regen2}

These discoveries constituted
a beautiful piece of quantum mechanics.
However, this was all immediately
overshadowed when $CP$-violation was observed
in the $K^0$ system \cite{cronin}.  Those authors were actually trying
to put a better limit on the fact that the $K_2$ did NOT decay
into $\pi^+\pi^-$.  But they found, at the level of $\sim 10^{-3}$,
that it did.
Therefore, the eigenstates of the complete weak system, including CP-violation,
are not quite $K_1$ and $K_2$, but rather a slightly different admixture
of  $K^0$  and $\bar{K^0}$.
These ``complete" eigenstates are called the $K_S$ and $K_L$  and are
parametrized by
\be
\ks = \frac{1}{[2(1+|\epsilon|^2]^(1/2}
      \bigg[ (1+\epsilon)\kz + (1-\epsilon)\kzb\bigg]~, \label{eqks}
\ee
\be
\kl = \frac{1}{[2(1+|\epsilon|^2]^(1/2}
      \bigg[ (1+\epsilon)\kz - (1-\epsilon)\kzb\bigg]~,  \label{eqkl}
\ee
$|\epsilon|$ being a number of order $10^{-3}$.

However,  $CP$-violation did allow one last
piece of quantum mechanics to be done.  It
meant that a direct interference measurement was possible because
both the $K_S$ and the $K_L$ decay into two-pion states.
A beam of $K^0$'s, which from Eqs. (\ref{eqks}) and (\ref{eqkl}) has an
equal amplitude ($a$) for being a $K_S$ and a $K_L$,
will decay into a $\pi^+\pi^-$ state with an intensity proportional to
\begin{eqnarray}
I_{\pi\pi}(\tau) &=& \frac{|\langle \pi^+\pi^-|K_0(\tau)\rangle|^2 }
                      {|P_{+-}|^2}\nonumber \\
&=& \exp[-\Gamma_S\tau]
   ~+~2|\eta_{+-}|~\exp[-(\Gamma_S + \Gamma_l)\tau/2]
           \cos(\Delta m\tau-\phi_{+-})  \nonumber \\
 &~& ~~~~~+~|\eta_{+-}|^2\exp[-\Gamma_L]~,  \label{pipi}
 \end{eqnarray}
 where
\be
P_{+-} = \langle\pi^+\pi^-\ks ~,
\ee
\be
\eta_{+-} = |\eta_{+-}|e^{i\phi_{+-}}=
\frac{\langle\pi^+\pi^-\kl}{\langle\pi^+\pi^-\ks}~,
\ee
\be
\Delta m = m_L-m_S~.
\ee
This prediction was verified \cite{Ktime1}.  By 1974 an experiment with
6 million events was able to show the oscillation very clearly
\cite{Ktime2}.
Figure 1 is taken from that paper.

The current experimental numbers of the various parameters are \cite{pdg}
\be
\tau_S = 1/\Gamma_S = (0.8926\pm 0.0012)\times 10^{-8}~sec~,
\ee
\be
\tau_L = 1/\Gamma_L = (5.17\pm 0.04)\times 10^{-8}~sec~,
\ee
\be
\Delta m = (0.5333 \times 0.0027)\times 10^{10} \hbar/sec~,
\ee
\be
|\eta_{+-}| = (2.269 \pm 0.023)\times 10^{-3} ~\approx~ |\epsilon|
\ee
\be
\phi_{+-} = (44.3 \pm 0.8)^o~.
\ee

There are, of course, many excellent discussions of this
phenomenon \cite{puzzle,cprev}.  but
in recent times,  interest in the  CP-violation
of the $K$ system has led to less
interest in the interference  phenomenon of itself \cite{pr,rmp}.  This is
 understandable since, to this
day, CP-violation is not  understood from a
fundamental theory.

However, the situation is changing, since there are now two other mixing
situations which have become of experimental interest.  One is that
of neutrinos (which I shall return to in Sec. 5) and the
other is the $B$ system.
With
the discovery of higher-mass quarks, it became clear that mixing
could reappear in other quark sectors; in particular,
in the ``beauty" sector of the neutral $B$ mesons.  Since the planned
B-factory may allow the study of CP-violation in this new system,
interest has revived in the  quantum interference.


\section{The Paradox}

The above has led to the formulation of a ``paradox" whose resolution,
both theoretically as well as by examples of
experimental manifestations in other
fields, is the focal point of our discussion.

Consider the neutral $B^0$-$\bar B^0$-meson system, whose properties are
similar to those of the $K$-meson system discussed in the last section.
Ignoring possible CP-violation and differences in lifetimes, one
could describe the two
``beauty" eigenstates as combinations of the two mass eigenstates,
$\bl$ and
$\bh$, which have masses $M_L$ and $M_H$.  Specifically,
\be
\bz = \frac{1}{\sqrt 2}\bigg( \bl+ \bh \bigg)~,
\ee
\be
\bzb = \frac{1}{\sqrt 2}\bigg( \bl- \bh \bigg)~.
\ee

An argument
has been presented \cite{lipkin} that it is contradictory to
try to discuss a beam of $B$ particles oscillating where the components
have different energies.  The argument is that only by looking at the
positions  of the components as functions of time,
and interpreting them as momenta, can one properly
describe things.

In particular, consider a $B^0$ produced at $x = 0$
 in a state of
definite energy, $E$.  The momenta of the $B_L$ and $B_H$
components,
  $p_L$ and $p_H$,   are given by
\be
p_L^2 = E^2 - M_L^2~, ~~~~~ p_H^2 = E^2 - M_H^2  ~.
\ee
Then, {\it as a function of} $x$, $B^0(x)$ will have a relative
mixture of $\bar B^0$ to $B^0$ of
\begin{eqnarray}
\left |
\frac{ \langle \bar B^0|B^0(x)\rangle} { \langle B^0|B^0(x)\rangle}
\right |^2
&=&
\left |
{{e^{ip_L x} - e^{ip_H x}}\over {e^{ip_L x} + e^{ip_H x}}}
\right |^2
=
\tan^2 \left({{(p_L - p_H)x}\over{2}}\right) \nonumber \\
&=&
\tan^2 \left({{(M_L^2 - M_H^2)x}\over{2(p_L + p_H)}}\right)~. \label{good1}
\end{eqnarray}
This is the normal $B - \bar B$ oscillation result.

Next the discussion  \cite{lipkin} considers
the case where a $B^0$ is produced
at time $t=0$ in a state of definite momentum, $p$ .
The energies of the
$B_L$ and $B_H$ components,
$E_L$ and $E_H$,  are
\be
E_L^2 = p^2 + M_L^2~,  ~ ~ ~ ~ ~ ~ E_H^2 = p^2 + M_H^2~.
\ee
Then, as a function of time,  $|B^0(t)\rangle$
will have relative components of $\bzb$ to $bz$ given by
\begin{eqnarray}
 \left | {\langle \bar B^0 |B^0(t)\rangle}\over
{ \langle B^0 |B^0(t)\rangle} \right |^2
&=&
\left |{{e^{iE_L t} - e^{iE_H t}}\over {e^{iE_L t} + e^{iE_H t}}}\right |^2=
\tan^2 \left({{(E_L - E_H)t}\over{2}}\right) \nonumber \\
&=&
\tan^2 \left({{(M_L^2 - M_H^2)t}\over{2(E_L + E_H)}}\right) \\
&\approx&
\tan^2 \left({{(M_L^2 - M_H^2)x}\over{4p}}\right)~, \label{good2}
\end{eqnarray}
where the last approximate equality is obtained by using
\be
x = vt = \frac{p}{E}\cdot t~.
\ee
Eqs. (\ref{good1}) and (\ref{good2}) are the same result,  so there
seems to be no problem.

Contrariwise,  the argument is raised \cite{lipkin}, if the states have
different momenta do they not also have different velocities,
 $v_L$ and $v_H$, so  that they therefore arrive at the point
x at different times, $t_L$ and $t_H$?
\be
x = v_Lt_L = {{p}\over{E_L}}\cdot t_L
= v_Ht_H = {{p}\over{E_H}}\cdot t_H ~.
\ee
Then  the
time-dependence of $\bz$ and $\bzb$  as a function of $x$ would be
\begin{eqnarray}
\left |{{ \langle \bar B^0|{B^o(x)\rangle}}\over
{ \langle B^0|{B^(x)}\rangle}}\right |^2
&=&
\left |{{e^{iE_L t_L} - e^{iE_H t_H}}\over {e^{iE_L t_L} +
e^{iE_H t_L}}}\right |^2=
\tan^2 \left({{(E_Lt_L - E_Ht_H)}\over{2}}\right) \nonumber \\
&=&
\tan^2 \left({{(M_L^2 - M_H^2)x}\over{2p}}\right)~. \label{bad}
\end{eqnarray}

The comparison of Eq. (\ref{bad}) with Eq. (\ref{good2}) is the
paradox.  There appears to be an inconsistency in how to make
a superposition
of different energy eigenstates.  One might even argue that
this ambiguity implies that one should not consider interference
between states of different energies.  Since this is commonly done
in the $K$ system by going to the center-of-mass system and
using the proper time, $\tau$, one would have to further explain
why a standard use of special relativity is not valid.

To understand
 the resolution of this paradox, one can return to the classic
quantum-mechanical interference problem, the double slit experiment.
If you know which slit the electron goes through, then you lose the
interference pattern.  If you know what arm of an interferometer the
``particle" goes through, you lose the interference.  This last has
been shown to be true even in ``delayed choice" experiments, where
the decision to find out which arm the particle is in is made {\it
after} the particle has entered the interferometer \cite{delay}.

The same is true here.  With interference, these are {\it not}
individual particles.  They are components in a mixed state.
If you know {\it which} of the two pure states,
$\bl$ or $\bh$,  you
have, then you lose the interference and the interference pattern
disappears.   In Sec. 5 we will point out where this result has
been made mathematically precise in the context of massive neutrino
propagation.


\section{Quantum Beats}

At this point it is
illuminating to show cases where
there is well-known and understood interference of states with different
energies.

A very clear example is that of ``quantum
beats," in atomic atomic and molecular physics.  As reviewed in
Refs. \cite{qbeat1,qbeat2,qbeat3}, quantum beats were first demonstrated
in 1964 without the use of lasers.  Since then, the use of lasers has
allowed the detection of quantum beats in Zeeman and hyperfine structures
of many atoms and molecules.
The example I give here is from a molecule, $S_1$ propynal
($HC\equiv CCHO$), because it
exhibits an interference structure  analogous to the $K$-meson system.

Consider the four-level system shown in Figure 2.  From the ground state
$|g\rangle$ a pair of closely-spaced excited states $|a\rangle$ and
$|b\rangle$ are excited by a short laser pulse of appropriate frequency
and bandwidth larger than the energy splitting of the excited states.
The coherent superposition of the two excited states at $t=0$, when the
laser stops, is given by
\be
|\psi(t=0)\rangle = \mu_{ga}|a\rangle + \mu_{gb}|b\rangle~,
\ee
where $\mu$ denotes a dipole matrix element:
\be
\mu_{jk} = \langle k|d|j\rangle~.
\ee
The time development of this state is now given by the frequencies
\be
\omega_a = E_{ga}/\hbar~, ~~~~\omega_b = E_{gb}/\hbar~,
\ee
and the decay constants $\gamma_a$ and $\gamma_b$:
\be
|\psi(t)\rangle  = \mu_{ga}\exp[-(\omega_a+\gamma_a/2)t]|a\rangle
          + \mu_{gb}\exp[-(\omega_b+\gamma_b/2)t]|b\rangle~.
\ee
Then the intensity of photons to the final state $|f\rangle$
will be proportional to
\begin{eqnarray}
I_d(t) &=& |\langle f|d|\psi(t)\rangle|^2 \\
       &=& |\mu_{ag}|^2 |\mu_{fa}|^2 \exp[-\gamma_a t]
           + |\mu_{bg}|^2 |\mu_{fb}|^2 \exp[-\gamma_b t] \nonumber \\
       &~& ~~~~ 2|\mu_{ag}\mu_{fa}\mu_{bg}\mu_{fb}|
      \exp[-(\gamma_a + \gamma_b)t/2]\cos[(\omega_a-\omega_b)t + \theta]~.
      \label{qbeat}
\end{eqnarray}
In Figure 3 we show the results from an experiment using $S_1$
propynal \cite{qbeat1,qbeat3}.  The main oscillation follows that
of Eq. (\ref{qbeat}), which exhibits the form of Eq. (\ref{pipi}).

In fact, there is  another  aspect to this system that is quite cute.
In actuality, this systems
is composed of two sets of two coherently excited levels, with quantum
beat frequencies of 16.6 MHz and 16.8 MHz.  Therefore, there is a
"beat" of the "beat frequencies," at 200 kHz.  That is seen in
the long period oscillation which occurs at about 2 $\mu$s in Figure 3.

But the point to be made is that these are different energy states
clearly exhibiting quantum interference when the position and
momentum of the states has nothing to do with the description.


\section{Other Phenomena}

\subsection{Rydberg wave packets}

During the past decade, a very interesting phenomenon has been
studied, Rydberg wave packets  \cite{ryd1,ryd2,ryd3,ryd4}.
A short-pulsed laser beam is used to excite a mixed state with
high-$\langle n
\rangle$.  This packet  has a significant overlap with a
number of  eigenstates of
different energies (and in fact is a squeezed state \cite{nie}).  That
these packets  exhibit classical motion and follow a classical
Kepler orbit is deduced from the following argument:

One measures the rate at which
the atoms decay from their excited energy packets.  There is an oscillation
in the
number of decays per unit time, and the oscillation period is that
which would obtain
for an elliptical Kepler orbit of that energy.
  If the particle is in an elliptical Kepler
orbit, then it would  undergo more acceleration near the perigee of
the orbit vs. near
the apogee.  Since the rate of decay increases with acceleration,
one infers that the oscillations reflect a coherent wave-packet
being near the perigee (more decays) and then heading towards the
apogee (fewer decays).

Here is an example of a  coherent wave-packet composed of
eigenstates of many energies.  It has a spatial coherence and  follows
a "classical-like" Kepler orbit.

\subsection{Neutron interferometry and the COW experiment}

Another interesting experiment was the COW experiment \cite{cow}.
Here individual
neutrons were sent through a single-crystal neutron interferometer.
The interferometer could be rotated about an axis defined by
the incoming neutron beam.  Therefore, upon entering the interferometer,
one part of the neutron beam would rise to a higher position
(and hence gravitational
potential) than the other.  The two split beams would propagate,
and then at the end the
second beam would be brought up to interfere with the first.
Depending upon the height difference, the interference-pattern fringes
shifted and hence yielded a measurement of local $g$ on the neutrons.

The interesting thing for us
here is that a complete analysis can and has been
given by Greenberger and Overhauser, using both the spatial and time
dependencies,  in the frameworks of
Galilean relativity, special relativity, and
general relativity, and in the lab and  freely falling frames \cite{go}.
There are no inconsistencies in the interpretations and they
agree with the experimental results.


\section{Neutrino Oscillations}

The last topic, that of neutrino oscillations, is the most timely.
They may have been observed
at the LAMPF beam stop.  The LSND group \cite{gorp} has
just reported the observation of electron-antineutrino events
 from a beam of incident muon antineutrinos obtained
by the decay of first $\pi^+$'s and then $\mu^+$'s.
This may explain the paucity of neutrrinos in solar neutrino
experiments \cite{davis}.  There, for example,
one might see only half the expected electron
neutrinos because the
other half are muon neutrinos by the time they reach the earth.

This puts even more interest in facilities like the proposed
Super Nova Burst Observatory \cite{snbo}.  This is a proposed underground
laboratory at the WIPP site in New Mexico.  Neutron counters would
observe the signal from all flavors of incident neutrinos (including
muon and tau neutrinos).  They would be produced by the neutral-current
interaction on nuclei.   The ordinary charged-current interaction of the
electron neutrinos would give a ``massless" time-of-arrival signal at other
laboratories, such as Kamiokande.  If the other neutrinos had a
significant non-zero mass, then their times-of-arrivals would
be delayed by the factor
\be
\delta t = 5.14 \times 10^{-3} R_{kpc} \left[\left(\frac{m_1}{E_1}\right)^2
              - \left(\frac{m_2}{E_2}\right)^2\right]~,
\ee
where $R_{kpc}$ is the distance to the supernova in kiloparsecs and
the $m$'s and $E$'s are the masses and energies of two distinct
neutrino species,

But this brings us back to our paradox, where the two $B$ mesons were
interfering and yet had distinct arrival times.  Kayer has analyzed the
massive neutrino scenario
exhaustively \cite{kayser}.  He showed that, if one takes into
account the fact that the beam is composed of wave packets of finite
size, not infinite plane waves,  when the wave packets no longer
overlap one can determine the individual arrival times and the
interference pattern is gone.


\section{Conclusion}

To summarize, quantum mechanics gives us a choice.  We can observe
interference or we  can tell which path
or where or what particle we have.  But if we have the latter,
then the interference is gone.

\section*{Acknowledgements}

I first wish to  thank Terry Goldman, who originally brought the paradox to
my attention and suggested there had to be a resolution.
I also thank Matthias Freyberger and J. Robert Huber
 for very helpful discussions.
This work was done
while visiting the University of Ulm under the aegis of the Alexander von
Humboldt Stiftung and with the support of the U.S. Department of Energy.



\newpage


\section*{Figure Captions}

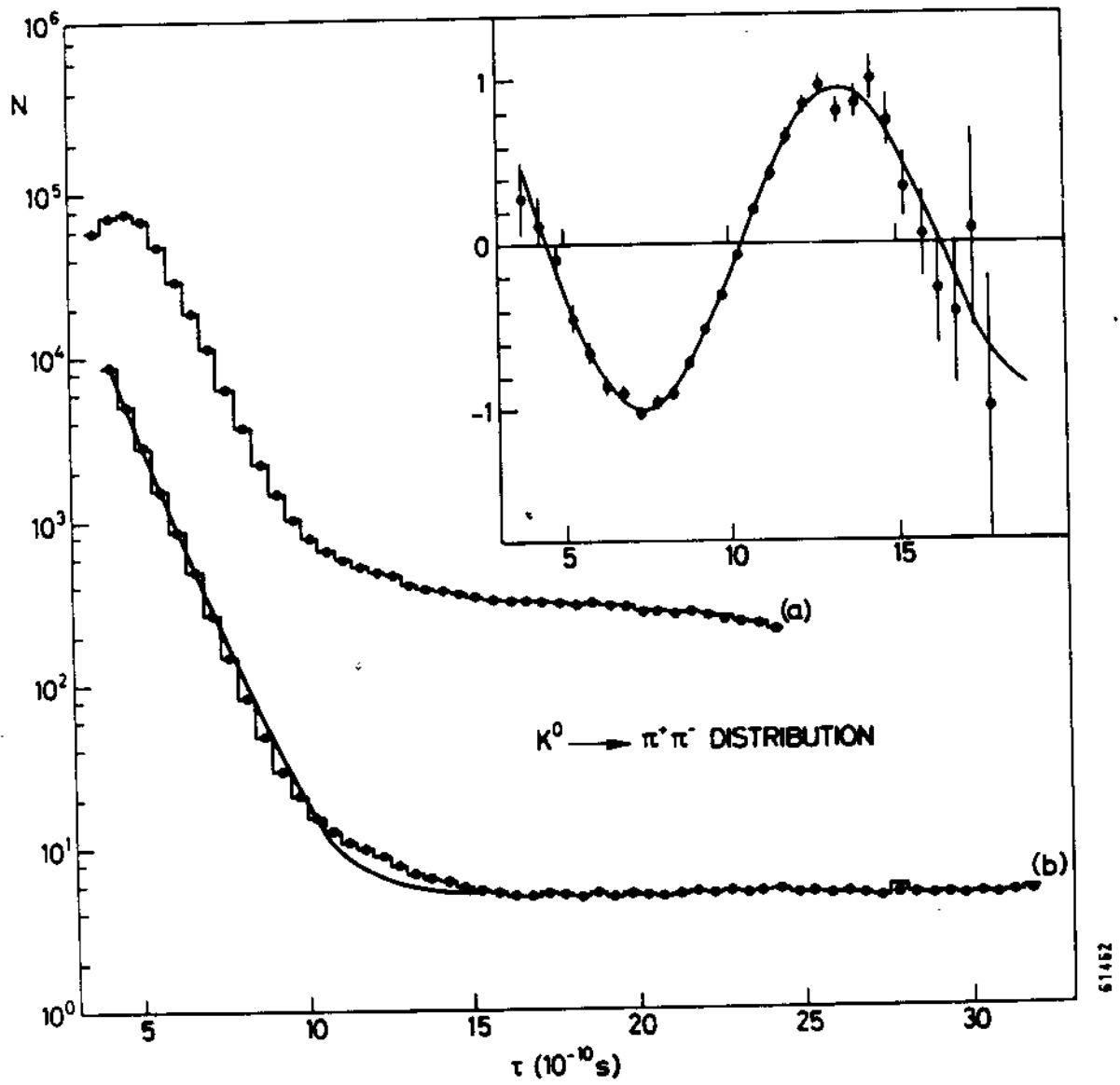
FIGURE 1.  Taken from Figure 4 of Ref. \cite{Ktime2}.  It shows the
time distribution of $ K \rightarrow \pi^+\pi^-$ events.  (a)
Events ({\it histogram}) and fitted distribution ({\it dots}).
(b) Events corrected for detection efficiency ({\it histogram}),
fitted distribution with interference term ({\it dots}), and
without interference term ({\it curve}).
Insert:  Interference term as extracted from data
({\it dots}) and fitted term ({\it line}). \\
{}~\\
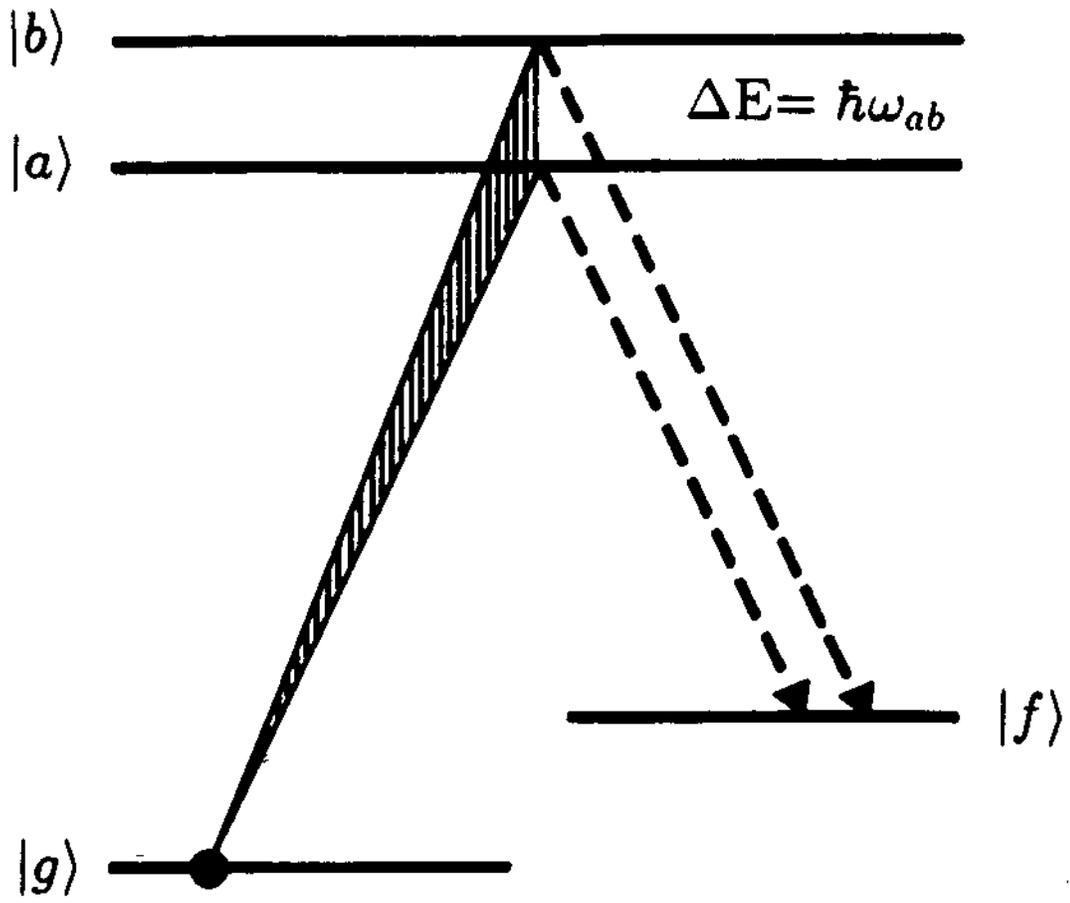
FIGURE 2.  Taken from Figure 1 of Ref. \cite{qbeat1}.
A four-level system.  .  States $|a\rangle$ and $|b\rangle$ are
coherently excited form a single ground state $|g\rangle$.
The coherence is evidenced by an interference effect (quantum beat)
when the emission decay to a common final state $|f\rangle$ is
observed. \\
{}~\\
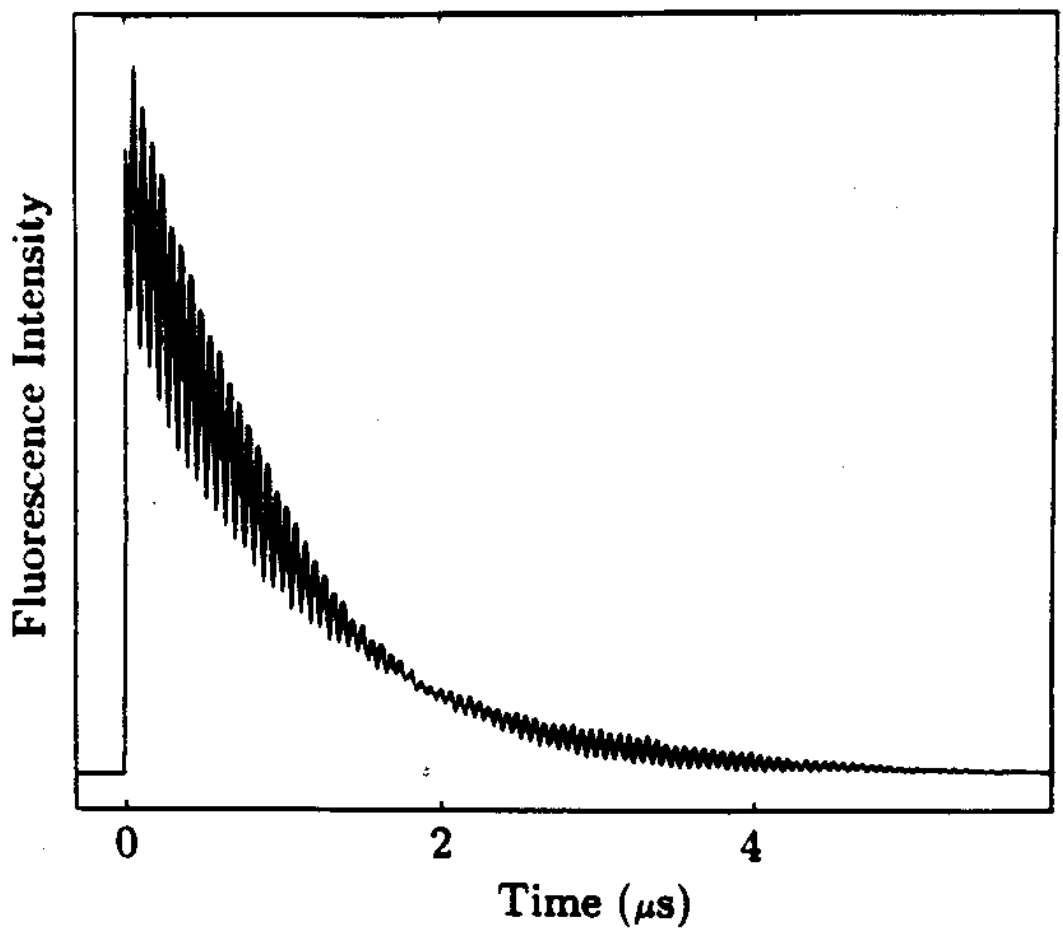
FIGURE 3.  Taken from Figure 2a of Ref. \cite{qbeat1}.
Quantum beats in the florescence decay of the $S_0^1$ band of
$S_1$ propynal.  Shown is the time domain signal.



\newpage

\begin{thebibliography}{99}


\bibitem{cpt} G. L\"uders,  Ann. Phys. (NY) 2 (1957) 1

\bibitem{gm} M.  Gell-Mann and A. Pais,  Phys. Rev. 97 (1955) 1387

\bibitem{3pi} K. Lande, L. M.  Lederman,  and W.  Chinowsky,
Phys. Rev. 105 (1957) 1925

\bibitem{pais} A. Pais and O. Piccioni,  Phys. Rev. 100 (1955) 1487

\bibitem{good} M. L. Good,  Phys. Rev. 106 (1957) 591; ibid. 110
(1958) 550

\bibitem{regen1}  R. H. Good, R. P.  Matsen, F. , Muller,
O. Piccioni, W. M.  Powell, H. S. White,
W. B.  Fowler, and R. W.  Birge,
Phys. Rev. 124 (1961) 1223

\bibitem{regen2} See also, e.g., the discussion in V. L. Fitch,
{\ Proceedings of the XIII International Conference on High Energy
Physics}, ed. by M. Alston-Garnjost (University of California Press, Berkeley,
1967),
p. 63

\bibitem{cronin} J. H. Christensen, J. W. Cronin, V. L. Fitch, and
R. Turlay,  Phys. Rev. Lett. 13 (1964) 138

\bibitem{Ktime1} M. Bott-Bodenhausen, X. De Bouard, D. G. Cassel,
D. Dekkers, R.  Felst, R. Hermod, I. Savin, P. Scharff,
M. Vivargent, R. R. Willitts, and K. Winter,
Phys. Lett. 20 (1966) 212

\bibitem{Ktime2} C. Geweniger, S. Gjesdal, G. Presser, P. Steffen,
J. Steinberger, F. Vannucci, H. Wahl, F. Eiself, H. Filthuth,
K. Kleinknecht, V. L\"uth, and G. Zech,
Phys. Lett. B 48 (1974) 487

\bibitem{pdg} Particle Data Group, Phys. Rev. D 50 (1994) 1173

\bibitem{puzzle} P. K.  Kabir, {\it The CP Puzzle} (Academic, London,
1968)

\bibitem{cprev} K. Kleinknecht,  Ann. Rev. Nucl. Sci. 26 (1976) 1

\bibitem{pr} R. Battiston, D. Cocolicchio, G. I. Pogli,
and N. Paver, Phys. Rep. 214 (1992) 293

\bibitem{rmp} B. Winstein and L. Wolfenstein,
 Rev. Mod. Phys. 65 (1993) 1113

\bibitem{lipkin}  H. J. Lipkin,  eprint hep-ph/9501269

\bibitem{delay} T. Hellmuth, H.  Walther, A. Zajonc, and
W. Schleich,  Phys. Rev. A 35 (1987) 2532

\bibitem{qbeat1} H. Bitto and J. R. Huber,  Opt. Comm. 80 (1990) 184

\bibitem{qbeat2} E. Hack and J. R. Huber,  Int. Rev. Physical Chem.
10 (1991) 287

\bibitem{qbeat3} H. Bitto and J. R. Huber,  Accounts Chem. Res. 25
(1992) 65

\bibitem{ryd1}  J. Parker and C. R.   Stroud, Jr.,
   Phys. Rev. Lett.
56 (1986) 716\\
J. A. Yeazell,  M.   Mallalieu, and C. R.  Stroud, Jr.,
  Phys. Rev. Lett.
64 (1990) 2007\\
J. A. Yeazell, J. A. and C. R.  Stroud, Jr.,
Phys. Rev. A 43 (1991) 5153

\bibitem{ryd2} A. ten Wolde, A., L. D.  Noordam, A. Lagendijk,
and H. B.  van
Linden van den Heuvell,   Phys. Rev. Lett. 61 (1988) 2099 \\
A. ten Wolde, L. D.   Noordam, A.  Lagendijk, and H. B.  van
Linden van den Heuvell,  Phys. Rev. A 40 (1989) 485

\bibitem{ryd3}   I. Sh. Averbukh and N. F.  Perelman,
  Phys. Lett. A 139 (1989)   449

\bibitem{ryd4}  G. Alber and P. Zoller,   Phys. Rep. 199
(1991) 231, and references therein

\bibitem{nie} M. M. Nieto,  Quantum Opt. 6 (1994) 9 \\
  The large round brackets in Eq. (5) should be squared.

\bibitem{cow} R. Colella, A. W.  Overhauser, and S. A.  Werver,
Phys. Rev. Lett. 34 (1975) 1472

\bibitem{go} D. M. Greenberger and A. W.  Overhauser,  Rev. Mod. Phys.
51 (1979) 43

\bibitem{gorp}
C. Athanassopoulos, L. B. Auerbach, R. Bolton, B. Boyd, R. L. Burman, D.
  O. Caldwell, I.Cohen, J. B. Donahue, A. M. Eisner, A. Fazely,F. J.
  Federspiel, G. T. Garvey, M. Gray, R. M. Gunasingha, V. Highland, R. Imlay,
  K. Johnston, W. C. Louis, A. Lu, J. Margulies, K. McIlhany, W. Metcalf, R. A.
  Reeder, V. Sandberg, M. Schillaci, D. Smith, I. Stancu, W. Strossman, G. J.
  VanDalen, J. N. BaW. Vernon, Y-X. Wang, D. H. White, D. Whitehouse, D. Works,
and Y. Xiao,
Phys. Rev. Lett. (submitted),
eprint nucl-ex/9504002

\bibitem{davis} R. Davis, Jr., B. T. Cleveland, and J. K. Rowley, in
{\it Science Underground}, AIP Conference Proceedings No. 96, ed. by
M. M. Nieto w. C. Haxton, C. M. Hoffman, K. W. Kolb, V. D. Sandberg,
and J. W. Toevs (AIP, New York, 1983), p. 2.

\bibitem{snbo}  D. B. Cline, G. M.  Fuller, W. P.  Hong, B. Meyer,
and J. Wilson,  Phys. Rev. D 50 (1994) 720

\bibitem{kayser} B. Kayser,  Phys. Rev. D 24 (1981) 110

\end{thebibliography}
\end{document}